\documentclass[11pt]{article}

\usepackage{fullpage}

\usepackage{booktabs} 

\usepackage{graphicx}
\usepackage[tight]{subfigure}

\usepackage{amssymb}

\usepackage{amsmath}
\usepackage{tcolorbox}

\usepackage{amsthm,tcolorbox}

\usepackage{ifthen}

\newtheoremstyle{stmt}
 {3pt}
 {3pt}
 {\itshape}
 {}
 {\bfseries}
 {}
 {1em}
 {(\thmnumber{#2})}

\theoremstyle{stmt}
\newtheorem{stmt}{}[section]
\newtheorem{stmt*}[stmt]{}

\tcolorboxenvironment{stmt*}{
  frame empty,
  colback=black!15!white,
  grow to left by=6pt,
  grow to right by=6pt,
  left=1.6pt,
  right=1.6pt,
  arc=0pt
}

\begin{document}

\newcommand{\pgraf}[3]{
\ifthenelse
 {\equal{#1}{h}}
  {\includegraphics[height=#2\textheight]{figs/#3}}
  {\includegraphics[width=#2\textwidth]{figs/#3}}
} 

\newcommand{\igraf}[3]{
\ifthenelse
 {\equal{#1}{h}}
  {\includegraphics[height=#2\textheight]{#3}}
  {\includegraphics[width=#2\textwidth]{#3}}
} 

\newcommand{\graf}[5]{ 
\begin{figure}[t]
\begin{center}
\igraf{#1}{#2}{#3}
\caption{{#5}\label{#4}}
\end{center}
\end{figure}
}

\newcommand{\agraf}[5]{ 
\begin{figure}[p]
\begin{center}
\igraf{#1}{#2}{#3}
\caption{{#5}\label{#4}}
\end{center}
\end{figure}
}

\def\endpf{\hfill $\Box$}

\newcommand{\prevs}[2]{
{\vskip 0.1in \noindent {\em Proof of \rf{#1}.} {#2} 
\vskip \belowdisplayskip}
}

\def\eps{{\varepsilon}}

\def\R{{\bf R}}

\renewcommand\^[1]{^{\langle#1\rangle}}
\renewcommand\-[1]{^{(#1)}}

\newcommand{\xhdr}[1]{\paragraph{\bf #1.}}

\newcommand{\omt}[1]{}

\newcommand{\rf}[1]{(\ref{#1})}

\newcommand\supproof[1]{}

\def\univ{U}
\def\coll{{\cal C}}
\def\seen{S}
\def\fint{t^*}
\def\zt{t^+}
\def\initmt{m_t^{(0)}}
\def\fincoll{{\cal F}}
\def\trueL{{K}}
\def\indset{A}
\def\allint{D}
\def\out{a}
\def\cat{\cdot}
\def\convset{B}

\newcommand{\clos}[1]{\langle {#1} \rangle}
\newcommand{\seq}[1]{\langle {#1} \rangle}

\newcommand{\res}[2]{{#1}_{#2}}

\newcommand{\gen}[2]{G_{#1}({#2})}

\newcommand\mxw[1]{r_{#1}}

\title{Language Generation in the Limit}

\author{
Jon Kleinberg\thanks{Departments of Computer Science and Information Science, Cornell University, Ithaca NY} \and
Sendhil Mullainathan\thanks{Booth School of Business, University of Chicago, Chicago IL}
}

\date{April 2024}

\maketitle

\begin{abstract}
Although current large language models are complex, the most basic specifications 
of the underlying language generation problem itself are simple to state:
given a finite set of training samples from an unknown language,
produce valid new strings from the language that don't already appear in
the training data.
Here we ask what we can conclude about language generation using
only this specification, without further assumptions.
In particular, suppose that an adversary enumerates the strings of an unknown
target language $L$ that is known only to come from one of a possibly infinite
list of candidates.
A computational agent is trying to learn
to generate from this language;
we say that the agent {\em generates from $L$ in the limit} if after
some finite point in the enumeration of $L$,
the agent is able to produce new elements that come exclusively from $L$
and that have not yet been presented by the adversary.
Our main result is that there is an agent that is able to
generate in the limit for every countable list of candidate languages.
This contrasts dramatically with negative results due to
Gold and Angluin in a well-studied model of language learning
where the goal is to identify an unknown language from samples;
the difference between these results suggests that 
identifying a language is a fundamentally different problem than
generating from it.
\end{abstract}

\section{Introduction}
\label{sec:intro}

The recent advances in large language models (LLMs)
have been remarkable, sparking active lines of
theoretical work into their performance.
These investigations implicitly revolve around
two fundamental questions: 
how do we formally reason about the effectiveness of LLMs;
and within such a framework, 
what are the core mathematical ideas 
that enable their performance?

Answers to these questions must begin by formalizing the specification
for what a generative algorithm for language should be doing.
Here, we propose starting from a very basic, assumption-free, statement for
such a specification:
there is an unknown target language $L$, 
over time the algorithm sees a sequence of strings
from $L$, and eventually we would like the algorithm to generate new strings
from $L$ that it has not seen before.\footnote{We will formalize the
definition of a language more precisely below, 
but for now we can think of a language as simply any
set of strings over a fixed alphabet; for example, the strings of the language
could be the set of all grammatical sentences (or all well-formed expressions)
according to a given grammar.}

Viewed this way, it is also clear why it seems so remarkable for
LLMs to be doing well at such a problem. 
The fully general statement of the problem feels unsolvable: 
if we know nothing about the unknown target language $L$, then how can a
generative algorithm reliably produce valid strings from $L$
that it hasn't seen before?

\xhdr{Language Learning in the Limit}
In fact, there is a well-established formalism that allows us
to phrase this point
precisely: the classical model of language learning in the limit, formulated
by Mark Gold in 1967 and fully characterized by Dana Angluin in 1980
\cite{gold1967language,angluin1980inductive}.
In this model, there is an unknown language $\trueL$ 
that is known only to be produced
by one of a list of candidate representations $R_1, R_2, R_3, \ldots$,
where $R_i$ produces some language $L_i$.
We can think of this list of representations as the set of all possible
context-free grammars, or the set of all possible finite automata, or the
set of all Turing machines with a fixed space bound, or any
other generative model that produces strings; in fact, the formal result
is much more general than this, in that it is sufficient to suppose that 
the unknown language $\trueL$ simply comes from a 
countable list of candidate languages
$L_1, L_2, L_3, \ldots$, and we can dispense with explicit representations 
altogether.\footnote{In this more general view, we will assume that the family
of languages is presented simply
via a black box that for a string $w$ and an index $i$
can answer the question, ``Is $w \in L_i?$''}

In the Gold-Angluin model, 
an adversary enumerates the strings of $\trueL$ one by one,
and the algorithm is required after each new string to guess a language $L_i$
from the list such that $L_i = \trueL$.
If there is some finite step $t$ after which the algorithm's guess is always
correct, then we say the algorithm has {\em identified $\trueL$ in the limit}.
Gold proved that this kind of language identification in the limit
is impossible in general, even for simple language
families such as the regular languages (i.e. those produced by finite automata),
and Angluin characterized
precisely those families for which it is possible, further establishing
how limited they are \cite{angluin1979finding,angluin1980inductive}.
Note, crucially, that in the Gold-Angluin model, the adversary enumerates
strings in $\trueL$, but does not provide examples of strings
that do not belong to
$\trueL$, nor does it allow the algorithm to ask questions about a string's
membership in $\trueL$;
their point with this formalism was to focus on cases where
an algorithm tries inferring a language purely from seeing a sufficient number
of examples of strings that belong to the language.

\xhdr{Our Results: Language Generation in the Limit}
These negative results of Gold and Angluin
feel intuitive --- how should we be able to identify
a language from a finite sample when we are allowed to make essentially
no assumptions about its structure?
Because of this intuition, both via the Gold-Angluin model 
and for more informal
reasons as well, the focus in language generation
has naturally turned to distributional assumptions;
one posits that large language models are so effective because they are able
to exploit distributional probabilities of language, and from a finite set of
samples they are able to estimate conditional probabilities of strings
with increasing accuracy.
In this way, the question moves from adversaries to probability distributions, 
and one seeks explanations for the effectiveness of LLMs through underlying
probabilistic models.

In this paper, we offer a sharply different view: we show that in the 
Gold-Angluin model of adversarially produced examples, 
{\em language generation is always possible.}
We will provide full details on the result and its proof 
beginning in the next section,
but the key point at a high level
is that even in an adversarial model with an unknown language $\trueL$, 
language generation is a fundamentally different task 
than language identification:
where identification asks an algorithm to eventually name a language
$L_i = \trueL$ after seeing a large enough finite sample $S$ from $\trueL$,
generation instead asks an algorithm to eventually output strings in 
$\trueL - S$ after seeing a large enough $S$ from $\trueL$.
Our main result is that this difference in specifications leads to dramatic
differences in what is possible in the limit; 
whereas the Gold-Angluin results establish that identification in the limit
is impossible except in highly constrained cases, 
we show that generation in the limit is possible for {\em every} countable 
list of candidate languages.

\xhdr{General Connections to Language Modeling}
Clearly, methods to design large language models in practice make extensive use
of the empirical distributional properties of language, as they should.
Our results don't question this design methodology;
when there are genuine empirical regularities in the training data,
there is no reason not to use them.
Rather, our results argue that if we are looking for the essential
reasons why language generation is tractable, 
we do not fundamentally require any empirical regularities, or indeed any
probabilistic assumptions at all; there is instead a formal sense in which 
language generation --- unlike broader
learning tasks such as language identification --- is possible even against
an adversary presenting positive training examples in a worst-case fashion.
In some crucial sense, the generation problem is therefore different from these
other learning tasks in ways that more detailed formalisms may
potentially obscure.

Despite the generality of the model, the generation algorithm
that proves our main theorem makes use of subtle structural
properties of the given list of candidate languages.
Again, we defer detailed descriptions to subsequent sections, but
the idea at a high level is to maintain a sequence of ``provisional languages''
that are consistent with the finite sample $S$
from $\trueL$ seen so far, and to continually refine this sequence of
provisional languages as the adversary adds strings to $S$.
Since the Gold-Angluin results say that the algorithm can never be sure
which is the true language $\trueL$, there is a sense in which 
this refinement process may need to continue
indefinitely, and in general it leads the algorithm to generate from
provisional languages that may be increasingly ``thin'' subsets of $\trueL$.
This does not cause trouble for the specification of language generation,
since it is acceptable to produce any unseen string from $\trueL$, but it does
mean that while the algorithm is able 
to eventually produce an infinite sequence of
unseen strings all from $\trueL$, 
it might do so from a narrow part of $\trueL$.

This property of the solution in the presence of an adversary suggests 
interesting connections to the problem of generation in practice as well.
First, and most directly, 
our basic model --- viewed at a high level given the abstract setting ---
is engaging in the basic loop that essentially
any method for language generation must perform:
iterating over a sequence of possible
representations for the target language, and 
repeatedly refining these representations
in the presence of new examples so as to
eventually achieve successful generation.

Beyond just this basic point, 
our model also encounters a key trade-off 
that appears in applications as well:
specifically, that
any method for generation has to deal with the
tension between an underlying
{\em validity problem} --- producing valid outputs --- and an underlying
{\em breadth problem} --- producing outputs that represent the full range
of valid outputs in some reasonable way.
The breadth problem is notoriously difficult, and it manifests itself
in numerous ways in the methodology of machine learning
and generative models.

The approach that proves our main result helps illustrate the tension between
validity and breadth even in settings with worst-case assumptions rather than
probabilistic ones, and this tension shows up in both the early
phases of our algorithm's execution and the later phases.
In the early phases, before the algorithm has refined its provisional
language sufficiently, it is generating too broadly and producing strings
that are not part of the target language $\trueL$ --- 
an analogue at a high level of a kind of 
hallucination in which the generated strings belong to some consistent
candidate language, but not to the actual target language
\cite{ji-hallucination-survey,kalai2023calibrated,xu2024hallucination}.
In the later phases, on the other hand, the algorithm continuously
shrinks its range of possible outputs so as to ensure that they
will be contained within $\trueL$ --- sacrificing validity for breadth
in a manner analogous to the issues that arise in the problem of
mode collapse for generative models 
\cite{arjovsky2017towards,arjovsky2017wasserstein}.
Our model therefore suggests interesting questions about the
fundamental trade-offs that may exist between validity and breadth 
even in settings without an underlying probabilistic model.

\section{Formal Model and Results}
\label{sec:model}

We now provide a formal description of the model and 
the statement of our results.
To begin with, we have a countable list of 
candidate languages $\coll = \{L_1, L_2, L_3, \ldots\}$, where each $L_i$
is a subset of some countable set $U$.
All we assume about the list of languages is that it is specified through
a black box that can answer questions of the form ``Is $w \in L_i$?'' for any
string $w \in U$ and language $L_i \in \coll$.
(If the reader finds it helpful for concreteness, 
they can consider the results that follow 
in the context of a specific list of languages $\coll$, such as the 
set of all context-free languages or the set of all regular languages;
but everything we say applies to general collections of languages.)
We will allow the collection $\coll$ to contain repetitions, in that 
we may have $L_i = L_j$ for different indices $i$ and $j$.
We will assume that all the languages $L_i$ are infinite; while the
original Gold-Angluin framework did not require this, it becomes important
in specifying the generation problem: if we require an algorithm to
output unseen strings forever, then this is not possible from a finite
language, where the algorithm would
eventually run out of new strings to generate.

An adversary and an algorithm now play the following game.
The adversary chooses a language $\trueL$ from $\coll$ without revealing it to the 
algorithm, and it begins enumerating the strings of $\trueL$ one by one over
a sequence of steps $t = 1, 2, 3, \ldots$.  The adversary can repeat strings
in its enumeration, but
the crucial point is that for every string $w \in \trueL$,
there must be at least one time step $t$ in which it appears.
Let $S_t$ be the set of strings that the adversary has enumerated in steps
1 through $t$.

\xhdr{Identification and Generation}
In this framework, we can now specify both the Gold-Angluin problem of
identification and the contrasting problem of generation 
that we study in this paper.
\begin{itemize}
\item {\em Identification (from \cite{gold1967language,angluin1980inductive}:} 
In each step, the algorithm observes $S_t$ and must
output an index $i$ (its guess for the true language $\trueL$).
The algorithm {\em identifies $\trueL$ in the limit} if there
is some $\fint$ such that for all steps $t \geq \fint$, the algorithm's guess
in step $t$ is an index $i$ for which $L_i = \trueL$.
\item {\em Generation (from the present paper):}
In each step, the algorithm observes $S_t$ and must
output a string $a_t$ (its guess for an unseen string in $\trueL$).
The algorithm {\em generates from $\trueL$ in the limit} if there
is some $\fint$ such that for all steps $t \geq \fint$, the algorithm's guess
$a_t$ belongs to $\trueL - S_t$.
\end{itemize}
Recall that 
a key point point about the Gold-Angluin framework is that the algorithm
is not provided with feedback about whether its outputs are correct ---
in the case of identification it is not told if its guesses about the identity
of the language are correct, and correspondingly our model of generation
does not provide the algorithm with feedback about whether the string
it generates in step $t$ belongs to the target language $\trueL$.

We know from the Gold-Angluin results that there is no algorithm that
can achieve identification in the limit for an arbitrary countable 
collection $\coll$
of languages (or even for specific countable collections, 
like the set of all regular
languages or the set of all context-free languages).
In contrast, our main result is a dramatically different answer for
language generation; it is possible for every countable collection:

\begin{stmt}
There is an algorithm with the property that for any countable
collection of languages $\coll = \{L_1, L_2, L_3, \ldots\}$,
and any enumeration of one of these languages $\trueL$,
the algorithm generates from $\trueL$ in the limit.
\label{stmt:main-infinite}
\end{stmt}

We prove \rf{stmt:main-infinite} in Sections \ref{sec:gen-f} and 
\ref{sec:gen-alg}.

\xhdr{A Result for Finite Collections}
We prove a second result as well, focusing on the variant of
the problem in which the collection of languages $\coll$ is finite.
In this case, it follows from Angluin's characterization that
every finite collection $\coll$ allows for identification in the limit.
Given this, what more might we ask for?
A natural question is whether there is a {\em uniform} bound on
the number of samples needed to ensure that the algorithm can correctly
identify the true language $\trueL$;
that is, for any finite collection $\coll$, is there a bound
$t(\coll)$ and an algorithm with the property that after seeing
any $t(\coll)$ distinct strings from $\trueL$, the algorithm is guaranteed to
correctly report $\trueL$ as its guess for the true language?

It is easy to see that for the Gold-Angluin model of language identification,
this is not possible.
For example, suppose that $\coll$
is the collection consisting of two languages $L_1$ and $L_2$:
$L_1$ consists of all possible strings, and
$L_2$ consists of all strings of even length.
Suppose there were a bound $t(\coll)$ and an algorithm that was guaranteed
to guess $\trueL$ correctly after seeing $t(\coll)$ distinct samples.
Then an adversary could present $t(\coll)$ distinct strings of even length,
and then ask the algorithm to guess whether the true language is $L_1$ or $L_2$:
if the algorithm guesses $L_2$ at this point, then the adversary
could announce that the answer is $L_1$, and conversely if the
algorithm guesses $L_1$.
This does not prevent the algorithm from learning the true language in the
limit, since the algorithm could simply keep guessing $L_2$ until
the first time (if ever) when a string of odd length is presented, 
at which point it switches to $L_1$.
But there is no fixed bound $t(\coll)$ by which it can be
guaranteed to output the correct guess.

However, for the problem of generation with a finite collection
of candidate languages, it is possible to
provide this much stronger type of uniform bound,
via an algorithm that can generate
correctly after seeing a finite sample $t(\coll)$ whose size is specified
at the outset.
In fact, we can achieve more: after seeing this finite sample,
the algorithm can correctly generate an infinite sequence of unseen elements
from the true language.

\begin{stmt}
There is an algorithm with the property that for any finite
collection of languages $\coll$, there is a number $t(\coll)$,
such that for any language $\trueL$ in $\coll$,
and any sequence $S$ of at least $t(\coll)$ distinct elements from $\trueL$,
the algorithm given $S$ can produce an infinite sequence of distinct strings
from $\trueL - S$.
\label{stmt:main-finite}
\end{stmt}

We prove \rf{stmt:main-finite} in Section \ref{sec:finite}.

\xhdr{Extensions and Generalizations}
Following these two main results in our basic model of generation, 
we provide (in Section \ref{sec:extend}) some extensions in a generalization
of the model.
Specifically, a familiar issue from language generation applications is the 
role of the {\em prompt}: a user provides an input string $p$,
and a generation algorithm is then supposed to produce a 
``continuation'' string $c$ to come after the prompt, so that
the concatenation of $p$ and $c$ is a valid utterance.
We offer a way of extending our model to incorporate the
notion of prompting, while maintaining the general structure of the model,
and we show how to formulate and prove a
generalization of our first main result in a setting where
at each time step the adversary is allowed to specify a prompt 
that must be completed.

\section{Initial Observations}
\label{sec:obs}

To begin with, we make a few additional notational points beyond
what was spelled out in the introduction. 
First, the languages $L_i$ we consider are all subsets of a countable
set of elements $U$, and for our two main results
\rf{stmt:main-infinite} and \rf{stmt:main-finite}, 
it is not important what $U$ corresponds to --- 
we can choose to think of $U$ as the natural numbers,
or the collection of all finite strings over a fixed alphabet, or any
other explicitly enumerated countable set.
In particular, it will not be crucial for these results whether
we talk about a set $U$ of strings listed as
$U = \{u_1, u_2, u_3, \ldots, u_m, u_{m+1}, \ldots\}$ or
simply the natural numbers 
$\{1, 2, 3, \ldots, m, m+1, \ldots\}$.
As a result, 
at different times in examples, for expositional simplicity
we will take $U$ to be different countable sets. 
(In contrast, when we consider extensions of our results to handle
constructions like prompts, 
it will be necessary to focus on the case in which $U$ is the
set of all strings over a finite alphabet, so that the notion
of string concatenation makes sense.)

Recall that we think of the algorithm as having knowledge of the sequence of
languages $\coll = \{L_1, L_2, L_3, \ldots\}$
in the following sense:
given an index $i$ and a string $w$, it can evaluate in finite time
whether or not $w \in L_i$.
We will refer to this evaluation of $w \in L_i$ as a
{\em membership query}, and assume henceforth that our algorithms
have the power to perform membership queries on the languages in $\coll$.
The true language $\trueL$ appears in the sequence
$\coll = \{L_1, L_2, L_3, \ldots\}$; suppose that $L_z = \trueL$
for an index $z$.
Going forward, we will often refer to $\trueL$ as 
$L_z$.
We observe that 
because languages can appear more than once in
the sequence $\coll$, it is possible that $\trueL$ is also equal
to $L_i$ for one or more indices $i \neq z$.
Finally, and crucially,
we note that the algorithm does not have the power
to pose queries about the membership of a string $w$ in the true
language $\trueL$: since 
the algorithm can only pose membership queries of the form ``$w \in L_i$?''
for strings $w$ and indices $i$ that it provides, it cannot ask
``$w \in \trueL$?'' because it does not know the index $z$ for which
$L_z = \trueL$.

We prove our main results beginning in Section \ref{sec:gen-f}, but first
we discuss a few points that provide useful background for
thinking about the problem.

\subsection{Review of Negative Results for Identification}
\label{subsec:gold}

The first point is a review of why language identification in the limit 
is not possible in general, adapting the exposition from
\cite{gold1967language,Lee:96a}.
It is useful to go through the proof of this result, so as
to get a better sense for the contrast with our positive results 
for generation.

There are many ways to show the negative result for language
identification using very simple language families, and
we choose one that highlights some intuitive contrasts with generation.
For the argument, let $U$ be the set of all integers, and let the collection
of languages $\coll$ --- each of which is a subset of $U$ --- be the set of all
infinite arithmetic progressions of integers.
(The choice of which countable ground set $U$ we use
is not crucial for any of these results,
and examples are often easier to describe over the set of integers than
over the set of finite strings.)
Formally, for an arbitrary integer $a$ and a positive integer $b$,
let $P_{a,b}$ be the arithmetic progression consisting of
all integers of the form $\{a + bi : i = 0, 1, 2, \ldots\}$;
and let $Q_{a,b}$ be the ``bidirectional'' arithmetic progression consisting of all
integers of the form $\{a + bi : i = \ldots, -2, -1, 0, 1, 2, \ldots\}$.
Let the collection $\coll$ consist of all arithmetic progressions
$P_{a,b}$ and $Q_{a,b}$.

Now, suppose by way of contradiction that there is an algorithm that can
identify in the limit
an adversarially chosen arithmetic progression $\trueL \in \coll$.
We construct an enumeration of $\trueL$ that causes the algorithm to fail, 
as follows.
First, for integers $a \leq b$, let $I[a,b]$ be the interval
of all integers $n$ for which $a \leq n \leq b$.
We enumerate elements of $U$ in stages, where each stage $s \geq 0$
consists of a set of consecutive steps.
If by induction stage $s$ has enumerated the elements of the
interval $I[-s,j(s)]$ for some $j(s) \geq 0$, then  
stage $s+1$ will enumerate additional elements so that by the end of
the stage we have enumerated exactly $I[-(s+1),j(s+1)]$
for some $j(s+1) > j(s)$.
In particular, stage $s+1$ first enumerates $-(s+1)$ and $j(s)+1$, and then it
begins enumerating $j(s) + 2, j(s) + 3, \ldots $ in increasing order.
At some point during this process, the algorithm must output 
$P_{-(s+1),1}$ as its guess for $\trueL$, since we can continue in this
way to produce a full enumeration of $P_{-(s+1),1}$, at which point the
true language is $\trueL = P_{-(s+1),1}$.
Once the algorithm outputs $P_{-(s+1),1}$ as its guess during stage $s+1$, 
we end stage $s+1$, defining $j(s+1)$ to be largest integer we've
enumerated up to that point, and we begin stage $s+2$.

In this way, the successive stages extend the interval $I[-s,j(s)]$
unboundedly in both directions.
We are therefore enumerating $Q_{0,1} = U$, and
so in fact $\trueL = Q_{0,1}$.
But we have also produced
an unbounded sequence of steps $t_0 < t_1 < t_2 < \cdots$
such that the algorithm guesses $P_{-j,1}$ at step $t_j$.
Thus, there is no time step $\fint$ for which
the algorithm outputs the (correct) guess $Q_{0,1}$ at every $t \geq \fint$.

\subsection{Generation and Closure}

In contrast, the particular collection $\coll$ defined above is easy for
an algorithm attempting to perform generation in the limit.
Once the algorithm has seen two elements $i < j$ from the true language
$\trueL$, then setting $b = j - i$, it knows that $\trueL$ must contain 
not just $i$ and $j = i + b$, but also $i + 2b, i + 3b, \ldots$
and in particular the entire arithmetic progression $P_{i,b}$.
Therefore, for every step $t$ from then on, it can always 
generate a string in $P_{i,b} - \seen_t$, since 
$P_{i,b}$ is infinite and $\seen_t$ is finite.
Given that $P_{i,b} \subseteq \trueL$, this is guaranteed
to be a string in $\trueL - \seen_t$.

A key point in this example is that the algorithm never needs to find
out the identity of the true language $\trueL$ in order to be able to
generate an infinite sequence of strings from it
(and by the argument above, we know that in fact it provably can't
find out the identity of the true language).
Rather, it just needs to answer the question, ``What is a string
that is guaranteed to belong to every language in $\coll$ consistent
with what I've seen so far?''

This highlights a sense in which generation is closely related
to a certain kind of {\em closure} operation on the languages in $\coll$,
which we define as follows.
Given the set $\seen_t$ of strings seen up to step $t$,
we say that a language $L_i \in \coll$ is {\em consistent} with $\seen_t$
if $\seen_t \subseteq L_i$.
We now define the {\em closure} of $\seen_t$ in $\coll$, denoted
$\clos{\seen_t}$, to be the intersection of all languages
in $\coll$ that are consistent with $\seen_t$.
We observe that $\seen_t \subseteq \clos{\seen_t}$, and
so the closure is non-empty.

Now, the more general point to take away from our example here
is that in any step $t$ where $\clos{\seen_t}$ contains
an element not in $\seen_t$, the algorithm can always safely output
an element in $\clos{\seen_t} - \seen_t$ and be sure that
it is outputting a new element in the true language $\trueL$.
This is simply because for every consistent language $L_i$,
$\clos{\seen_t} - \seen_t \subseteq L_i - \seen_t$
by the definition of the closure operation, and
in particular this holds for the true language $\trueL$.
This is what the algorithm did in our example:
once $\seen_t$ contained the elements $i$ and $i+b$, then
we could conclude that the full arithmetic progression
$P_{i,b}$ was a subset of $\clos{\seen_t}$ and would be for all 
future steps $t$, so this provided an infinite set that the algorithm
could safely output elements from.

We refer to $\clos{\seen_t}$ as the ``closure'' of $\seen_t$
by analogy with other forms of ``closure.''
For example, the convex hull of a set of points $S$ in the plane
follows a similar idea: it is simply the intersection of all convex sets
that contain $S$.
And although we won't attempt to define a formal contrast between
learning and generation for convex sets here, it is clear informally
that the contrast we've been discussing in this section has an
analogue in this geometric setting as well.
For example, suppose an adversary is thinking of a hidden convex set $\convset$,
and it shows a finite set $S$ of examples to an algorithm.
From any finite set, the algorithm has no chance of identifying the true
convex set that the adversary is thinking of; but the algorithm can
correctly generate an infinite sequence of new points from $\convset$ by simply
enumerating points in the convex hull of $S$ that do not already belong to $S$.

\xhdr{The insufficiency of closure}
Closure is a first useful idea in designing an algorithm that can generate
in the limit, and we will see
in Section \ref{sec:finite} that
for the case of finite collections $\coll$ it is
the main idea that we need.
Unfortunately, it is insufficient on its own to achieve generation
in the limit for arbitrary countable collections $\coll$,
for the simple reason that $\clos{\seen_t} - \seen_t$ can be empty
in general, providing the algorithm with no guidance on which element to
generate next.

To see how this can happen in a simple example, consider
the following slightly more complicated collection of languages $\coll$.
$U$ is still the integers, and now for every
arithmetic progression $P_{a,b}$ and every finite set of integers $V$,
we define the language $L(a,b,V) = P_{a,b} \cup V$.
Our collection $\coll$ consists of every $L(a,b,V)$ for 
an arbitrary integer $a$, an arbitrary positive integer $b$, and an arbitrary
finite set of integers $V$.
(Think of $L(a,b,V)$ as a copy of the arithmetic progression $P_{a,b}$
that has been obscured by an arbitrarily large finite set $V$
so that its structure is harder to discern.)
We could also include the languages $Q_{a,b} \cup V$ in $\coll$
for every integer $a$, positive integer $b$, and finite set $V$, 
but this won't be crucial for our purposes in this example.

Now, suppose the adversary is enumerating a language $\trueL = L(a,b,V)
\in \coll$,
and consider the set $\seen_t$ of samples after $t$ steps.
We claim that $\clos{\seen_t} = \seen_t$.
Intuitively, this is because $\seen_t$ might have come completely 
from the finite set $V$ that is part of $L(a,b,V)$;
more formally, there cannot be any element $j \in \clos{\seen_t} - \seen_t$
because $L(j+1,1,\seen_t)$ is consistent with $\seen_t$, and it does not
contains $j$.

This example illustrates the sense in which we mean that closure by
itself is insufficient to provide a strategy for generation in the limit;
in this instance, the algorithm has to repeatedly output elements outside
the closure of $\seen_t$ and yet somehow eventually generate from $\trueL$.
To see how it does so in this current example, suppose that in any step $t$,
the algorithm finds the two largest elements $i < j$ in $\seen_t$, and with
$b = j - i$, it outputs $i + 2b$.
The point is that if the true language $\trueL$ is $L(a,b,V)$,
there will come a time when the adversary has enumerated
every element of $V$, and within a finite number of steps after this,
the two largest elements of $\seen_t$ will have to come from 
$P_{a,b}$.  From this step onward, the algorithm's strategy of outputting
$i + 2b$ is guaranteed to produce an element in $\trueL - \seen_t$.
This particular strategy of course seems highly specific to 
the particular example we are discussing here, but at a very high
level it does contain ideas that are more easily discernable in
retrospect from the general solution, which we describe in the 
next two sections.

\subsection{The Failure of Direct Hypothesis Enumeration}

We now discuss a final point that is useful to 
highlight before turning to the main proof.
In thinking about approaches to generation in the limit, 
there is a natural strategy
that at first seems to solve the problem directly, but in fact does not work.
Its failure is useful to discuss, since it motivates the more involved solution
that follows.

The strategy is to move through the list of languages 
$\coll = \{L_1, L_2, L_3, \ldots\}$
in order, treating each language $L_i$ 
as a hypothesis for $\trueL$ until the sample
$S_t$ proves otherwise.
That is, we start with $L_1$, and we generate strings from $L_1 - S_t$ 
until we encounter (if ever) a step $t$ in which $S_t \not\subseteq L_1$.
At this point we know that $L_1$ cannot be the true language $\trueL$, and so we
continue the process with $L_2$.
The nice idea that underpins this strategy is that the true language $\trueL$ is
equal to $L_z$ for some index $z$.  
So if our process were to reach $L_z$ at some step $\fint$, 
it would never move on from $L_z$, and so we would be generating from 
$\trueL - S_t$ for all $t \geq \fint$.
(Since $\coll$ can contain repetitions,
$\trueL$ might appear several times, 
but we can take $L_z$ as the first appearance.)

Unfortunately, there is a deep problem with this approach:
there may be a language $L_i \in \coll$ with the property that
$L_i$ comes before $L_z$ and $L_i$ properly contains $L_z$
(that is, $i < z$, and $L_z \subsetneq L_i$).
In this case, our procedure would stop at the first such $L_i$ forever:
since it is only ever shown samples in $S_t$ that come from the language $L_z$,
and since $L_z \subseteq L_i$, it would never encounter a string in $S_t$
that didn't belong to $L_i$, and so it we would never move on to $L_{i+1}$.
And when this procedure generated from $L_i - S_t$, there is no guarantee that it
would choose strings from $L_z$.
(Recall that the algorithm is not provided with feedback about whether
the string it generates in step $t$ belongs to the target language $L_z$;
its only knowledge of $L_z$ comes from the sample $S_t$ that the adversary
is enumerating.)

This problem is not easily avoided, since if this approach worked as written, 
it would also solve identification in the limit, which we know is impossible.
So we need to extract some of the useful ideas from this failed approach ---
in particular, the point that $\trueL$ appears at some finite index in the 
list $\coll$, as the language $L_z$ ---
but add important further ideas as well.
Specifically, if the algorithm is maintaining hypotheses for the true language
$\trueL$ over time, it can provably never know whether its current hypothesis is 
correct; instead, it must be always moving further down the collection of
languages, potentially considering languages that are not $\trueL$, 
but in such a way
that it is eventually always generating from $\trueL - S_t$.
This is what our proof beginning in the next section will have to accomplish.

\section{Generation in the Limit via a Function}
\label{sec:gen-f}

We prove our main result in two parts.
We first give a method for language generation in the limit that
is not concerned with the computational power required by the
agent performing the generation.
Thus, rather than an algorithm to generate the string,
we ask whether we can construct a function $f_\coll$ based
on the given language collection that maps 
a finite set of strings to a new string;
this function takes the strings $S_t$ seen so far and outputs a
string $f_\coll(S_t)$ intended to be in $\trueL - S_t$.
We will prove the following:

\begin{stmt}
For every countable collection of languages $\coll$,
there is a function $f_\coll$ from finite subsets of $U$ to elements of $U$,
such that for every enumeration of a language $\trueL \in \coll$,
there is a $\fint$ such that for all $t \geq \fint$, we have
$f_\coll(\seen_t) \in \trueL - \seen_t$.
\label{stmt:main-noncomput}
\end{stmt}

Note that while this positive result is not concerned with
the computational power required to evaluate $f_\coll$, it already
contains the core contrast with language identification, which
remains impossible even if we simply ask for a function $f_\coll$,
by the same argument given in Section \ref{subsec:gold}.
In the next section, we will then go on to prove 
\rf{stmt:main-infinite} by using an algorithm that only performs
standard computational steps and membership queries 
of the form ``$w \in L_i$?''

\xhdr{Minimal and critical languages}
As before, we will suppose $z$ is an index such that $L_z = \trueL$.
We say that a language $L_i$ is {\em consistent} with the sample
at step $t$ if $\seen_t \subseteq L_i$.
An important idea, which is implicit in our discussion of
the failed approach at the end of Section \ref{sec:obs}, is that
if $L_i \subseteq L_j$ are both consistent with $\seen_t$, then
it is safer for an algorithm to generate from $L_i$ than from $L_j$:
if $w \in L_i - \seen_t$ then we must also have $w \in L_j - \seen_t$.
This suggests that it would be useful to find consistent languages
that are {\em minimal} with respect to inclusion:
we say that $L$ is {\em minimal} if 
$L \in \coll$ is consistent with $\seen_t$, and 
there is no $L' \neq L$ such that $L'$ is consistent with $\seen_t$
and $L' \subseteq L$.
Unfortunately, this is too much to ask for, since there exist
instances of the problem for which there might not be any
languages that are minimal with respect to inclusion.
(In a finite collection of language there would need to be a minimal language,
but it is easy to construct infinite collections without one.)

Therefore, we define a related concept that only involves
examining the inclusion of a given language with respect
to a finite set of other languages.
Specifically, we look for languages 
$L_n$ that are consistent with $\seen_t$ in a given step $t$,
such that $L_n$ is a subset of every consistent language 
that {\em precedes} it in the indexing of $\coll$.
We will say that such a language is {\em critical} at step $t$.
To define this formally, we first let
$\res{\coll}{n}$ denote the finite collection
of languages $\{L_1, L_2, \ldots, L_n\}$.
We now have the following definition.

\begin{stmt}
A language $L_n$ is {\em critical} at step $t$ if 
$L_n$ is consistent with $\seen_t$, and for every language 
$L_i \in \res{\coll}{n}$ that is consistent with $\seen_t$,
we have $L_n \subseteq L_i$.
\label{def:critical-def}
\end{stmt}

\xhdr{Finding critical languages}
At any given step $t$, there is 
at least one language consistent with $\seen_t$,
since the language $L_z = \trueL$ is always consistent with $\seen_t$.
It follows that there is also at least one critical language at any step $t$:
for any $t$, the consistent language $L_i$ with
the lowest index $i$ must be critical at step $t$,
as it is the only consistent language in $\res{\coll}{i}$.

Note that there can be choices of $t$ for which the language $L_z = \trueL$
is not critical at step $t$.
But a crucial fact is that $L_z$ will eventually become critical at some
step $t$ and remain critical forever after that:
We prove this next.

\begin{stmt}
There exists a time step $\zt$ such that for all $t \geq \zt$,
the language $L_z$ is critical at step $t$.
\label{stmt:true-critical}
\end{stmt}

\proof{
Let $\indset$ be the indices $i < z$ for which $L_z \not\subseteq L_i$.
For each $i \in \indset$, let $v_i$ be an element of $L_z - L_i$.
Let $t_i$ be the step in which $v_i$ first appears in the enumeration of $L_z$,
and let $\zt = \max_{i \in \indset} t_i$.

Now, suppose by way of contradiction that for some $t \geq \zt$, the
language $L_z$ is not critical at step $t$.
In this case, there must be some 
$L_i \in \res{\coll}{z}$
such that $L_i$ is consistent with $\seen_t$ and 
$L_z \not\subseteq L_i$.
But we know that $v_i \in \seen_t$ and $v_i \not\in L_i$,
contradicting the consistency of $L_i$ with $\seen_t$.
\endpf
}

There can be multiple critical languages at a given step $t$;
for example, if on the step $\zt$ in \rf{stmt:true-critical}
the first consistent language $L_i$ is not equal to $L_z$, then both
$L_i$ and $L_z$ will be critical at step $\zt$.\footnote{This is 
also a useful moment
to recall that there can multiple languages $L_i \in \coll$ that
are equal to $L_z$.
A direct analogue of the proof of \rf{stmt:true-critical}
shows that for any $i$ for which $L_i = L_z$, there is a step
$\zt_i$ such that for all $t \geq \zt_i$, the language $L_i$
is critical at step $t$.
But these steps $\zt_i$ may be different for different $i$.
In particular, if $i < j$ are indices with
the property that $L_i = L_j = L_z$, then we
must have $\zt_i \leq \zt_j$, but it is possible that $\zt_i < \zt_j$,
simply because at a given time step $t \geq \zt_i$, 
there might be a language $L_h$ with $i < h < j$ such
that $L_h$ is consistent with $\seen_t$ and $L_j \not\subseteq L_h$.
For our purposes in the arguments to follow, it is sufficient to
consider the step at which a specific copy of the true
language $L_z$ first becomes critical forever;
we do not need to
worry about whether other copies are critical in this step or not.}
Despite the potential multiplicity of critical languages,
the collection of
all critical languages at step $t$ has a useful structure
that follows directly from the definition of criticality.

\begin{stmt}
Let $i < j$, and suppose that $L_i$ and $L_j$ are both critical at step $t$.
Then $L_j \subseteq L_i$.
\label{stmt:critical-nested}
\end{stmt}

\proof{
$L_i$ belongs to $\res{\coll}{j}$ and is consistent with $\seen_t$, 
and $L_j$ is critical at step $t$,
so by definition \rf{def:critical-def},
we have $L_j \subseteq L_i$.
\endpf
}

\xhdr{A function for generation in the limit}
At a given step $t$, suppose that the critical languages are
$L_{n_1}, L_{n_2}, L_{n_3}, \ldots$ where
$n_1 < n_2 < n_3 < \cdots$.
(This list of critical languages might be finite or infinite.)
Then \rf{stmt:critical-nested} tells us that this sequence is nested
by inclusion:
$L_{n_1} \supseteq L_{n_2} \supseteq L_{n_3} \supseteq \cdots$.

By \rf{stmt:true-critical} we know that the language $L_z$
will eventually appear on this nested list from some step $\zt$ onward,
but even then 
we do not know which index $n_i$ it corresponds to 
at any given step $t \geq \zt$.
Indeed, to recall a point from earlier,
the Gold-Angluin results for learning in the limit tell us that
we can never know for sure which index corresponds to $L_z$.
But we now arrive at the crucial point, which is that beyond
some finite index, all the critical languages are subsets of $L_z$,
so it is safe to generate from any of them.

Given this, we are prepared to construct our function $f_\coll$.

\begin{stmt}
$f_\coll(\seen_t)$ is defined as follows.
We first identify all languages in $\res{\coll}{t}$ that are 
critical at step $t$.
(If no such languages exist --- which can only happen if none
of them are consistent with $\seen_t$ --- we 
define $f_\coll(\seen_t)$ arbitrarily.)
Among these critical languages, let $L_{n_t}$ be the one with the
largest index $n_t \leq t$.
We define $f_\coll(\seen_t)$ to be the lowest-indexed element of
$L_{n_t} - \seen_t$.
\label{def:f}
\end{stmt}

To prove our initial claim \rf{stmt:main-noncomput}, 
it is sufficient
to verify the following property of $f_\coll$.

\begin{stmt}
For any language $L_z \in \coll$ and any enumeration of $L_z$,
there is a $\fint$ such that for all $t \geq \fint$, we have
$f_\coll(\seen_t) \in L_z - \seen_t$.
\label{stmt:f-property}
\end{stmt}

\proof{
In the given enumeration of $L_z$, 
\rf{stmt:true-critical} tells us that there will come a step $\zt$
such that for all $t \geq \zt$, the language $L_z$ is critical at step $t$.
Let $\fint = \max(z,\zt)$.
In every step $t \geq \fint$, our construction of $f_\coll$ will
include $L_z$ among its critical languages in $\res{\coll}{t}$.
Therefore, 
the highest-indexed critical language $L_{n_t} \in \res{\coll}{t}$
satisfies $n_t \geq z$, and so
by \rf{stmt:critical-nested}
we have $L_{n_t} \subseteq L_z$.
Since $f_\coll(\seen_t) \in L_{n_t} - \seen_t$, we have
$f_\coll(\seen_t) \in L_z - \seen_t$ as required.
\endpf
}

As a final note, we observe that the current formulation
of $f_\coll$ allows it to generate the same string more than once,
provided that this string is in $\trueL - S_t$.
However, it is not hard to modify $f_\coll$ so that it generates
a different string each time, essentially by defining it so
that it generates the lowest-indexed element that it hasn't
already generated.

\xhdr{The computational power required to produce $f_\coll$}
Our plan was to construct $f_\coll$ without worrying about
the computational power required to do so (and recalling that
for comparison, 
in the corresponding problem of identification in the limit, no
function $f_\coll$ achieving identification 
could exist regardless of the computational
power required to produce it).
Now that we've constructed an appropriate $f_\coll$, we can ask
what was in fact required computationally.

In addition to standard computational steps and membership queries of the
form ``$w \in L_i$?'', the definition of $f_\coll(S_t)$
requires that we identify the critical languages in $\res{\coll}{t}$.
From the definition, 
we can do this provided we can
answer a finite number of {\em subset queries} 
of the form ``$L_i \subseteq L_j$?''.
So an algorithm augmented with the power to perform
such subset queries can perform generation in the limit.

In the next section, we will show how to remove the necessity
for subset queries, so that generation in the limit can be performed
by an algorithm using only standard computational steps and 
membership queries.

\section{Generation in the Limit via an Algorithm}
\label{sec:gen-alg}

We now prove \rf{stmt:main-infinite} by giving an algorithm
that generates in the limit for any countable collection of languages $\coll$,
using only standard computational steps and membership queries
of the form ``$w \in L_i$?''

The set of possible strings $U$ can be written as
$U = \{u_1, u_2, u_3, \ldots\}$, and for simplicity we will sometimes
use the language of the positive integers to describe $U$, treating
$u_i$ as the number $i$.
In an enumeration of the true language $L_z = \trueL$, let the sequence of
strings that are enumerated step by step 
be denoted $w_1, w_2, w_3, \ldots$.

\xhdr{Extending definitions to finite subsets of languages $L_i$}
The notion of a critical language was crucial to our approach in the
previous section, and since 
the direct approach to verifying whether a language is critical involves 
subset queries, an important part of designing an algorithm that avoids
subset queries is to work with finite subsets 
of the languages in $\coll$.
Thus, for a language $L_i \in \coll$ and a number $m$, we will use
$L_i[m]$ to denote the finite set $L_i \cap \{u_1, u_2, \ldots, u_m\}$.
Note that deciding whether $L_i[m] \subseteq L_j[m]$ for a fixed value of $m$
requires only that we make at most $2m$ membership queries: we simply ask
whether $u_h \in L_i$ implies $u_h \in L_j$ for all $h \leq m$.
This allows us to replace the definition of critical languages 
from \rf{def:critical-def}
with a variation tailored to finite sets, and to be able to verify that
this finite version is satisfied using only membership queries.
By gradually expanding the value of $m$ over the steps $t$ of the algorithm,
we will eventually get to large enough values of $m$ and $t$ for which
this finite version of the definition is sufficient for generation.

Our extension of definition \rf{def:critical-def} for critical languages
to finite sets is as follows.

\begin{stmt}
Let $t$ and $m$ be positive integers.
A language $L_n$ is 
$(t,m)$-critical if $L_n$ is consistent with $\seen_t$,
and for every language $L_i \in \res{\coll}{n}$ 
such that $L_i$ is consistent with $\seen_t$,
we have $L_n[m] \subseteq L_i[m]$.
\label{def:finite-critical}
\end{stmt}

Since $L_n \subseteq L_i$ implies 
$L_n[m] \subseteq L_i[m]$ for any $m \geq 1$, we have the following
analogue of \rf{stmt:true-critical}.

\begin{stmt}
There exists a time step $\zt$ such that for all $t \geq \zt$ and
all $m \geq 1$, 
the language $L_z$ is $(t,m)$-critical.
\label{stmt:finite-true-critical}
\end{stmt}

The analogue of \rf{stmt:critical-nested} also still holds with this
definition, using the same proof.

\begin{stmt}
Let $i < j$ and suppose that $L_i$ and $L_j$ are both $(t,m)$-critical.
Then $L_j[m] \subseteq L_i[m]$.
\label{stmt:finite-critical-nested}
\end{stmt}

Finally, there is a basic monotonicity property of
$(t,m)$-criticality that is useful to write down.

\begin{stmt}
Suppose that $L_n$ is $(t,m)$-critical, and $m' < m$.
Then $L_n$ is $(t,m')$-critical.
\label{stmt:finite-critical-monotone}
\end{stmt}

\proof{
Since $L_n$ is $(t,m)$-critical, we know it is consistent with $\seen_t$,
and that $L_n[m] \subseteq L_i[m]$ for all languages
$L_i \in \res{\coll}{n}$ such that $L_i$ is consistent with $\seen_t$.
Now, if $L_i$ is a language in $\res{\coll}{n}$ that is consistent with
$\seen_t$, then since $L_n[m] \subseteq L_i[m]$ and $m' < m$, we have
$L_n[m'] \subseteq L_i[m']$. 
It follows that $L_n[m'] \subseteq L_i[m']$ for all 
$L_i \in \res{\coll}{n}$ such that $L_i$ is consistent with $\seen_t$,
and so $L_n$ is $(t,m')$-critical.
\endpf
}

\subsection{An algorithm for generation in the limit}

We now describe an algorithm for generation in the limit.
As before, $S_t = \{w_1, w_2, \ldots, w_t\}$ 
is the subset of $\trueL$ enumerated through step $t$,
treating the $w_i$ as integers.
We will consider the languages in $\res{\coll}{t}$
in step $t$, and 
maintain an auxiliary variable $m_t$,
roughly corresponding to how large a prefix $L_i[m]$ we consider
from each language $L_i \in \res{\coll}{t}$.

At the start of step $t$, we set $\initmt = \max(m_{t-1}, w_t)$;
note that by induction this implies $\initmt \geq \max_{t' < t} w_{t'}$.
(At the end of step $t$, we will define $m_t$ to be a number that is
at least as large as $\initmt$, via the process described below.)
We then determine which $L_i \in \res{\coll}{t}$ are consistent
with $\seen_t$; note that by the definition of $\initmt$, it is sufficient
to perform membership queries for only the finite set of elements in
$L_i[\initmt]$ in order to do this.
If there are no consistent languages in $\res{\coll}{t}$, 
then we output a string arbitrarily.

Otherwise, there is at least one language consistent with $\seen_t$, 
and so there is at least one $(t,m)$-critical language for any choice of $m$,
since the first consistent language is $(t,m)$-critical for all $m$.
Our goal is to imitate the plan from \rf{def:f} and generate a new
string from the highest-indexed critical language.
But to do this, we have to find a new string, and this will in general 
require performing additional membership queries.

\xhdr{Generating a string}
For any choice of $m$, let $n_t(m)$ be the maximum index of
a $(t,m)$-critical language from $\res{\coll}{t}$;
as noted above, $n_t(m)$ is well-defined since we are in the case
where at least one language in $\res{\coll}{t}$ is consistent with
$\seen_t$, and so the first consistent 
language is $(t,m)$-critical for all $m$.
We now search for a string to generate as follows.

We maintain a counter $m$ that begins at $\initmt$ and gets incremented 
iteratively, with each iteration doing the following:
\begin{itemize}
\item[(i)] Increment $m$ by 1.
\item[(ii)] Perform membership queries to determine $L_i[m]$ for each 
$L_i \in \res{\coll}{t}$.
Note that since $m \geq \initmt \geq \max_{t' < t} w_{t'}$, 
the determination of which languages in $\res{\coll}{t}$ are consistent
with $\seen_t$ does not change (relative to the initial iteration) when we do this.
\item[(iii)] Determine which languages are $(t,m)$-critical, and from 
this determine $n_t(m)$.
Note that this only requires consulting the results of
membership queries already performed; also, since we are working with a
value of $t$ for which $\res{\coll}{t}$ contains at least one consistent
language, the value of $n_t(m)$ is well-defined.
\item[(iv)] If there is any string $u_i$ for $i \leq m$ such that
$u_i \in L_{n_t(m)} - S_t$, then 
choose the minimum $i$ with this property;
output the string $u_i$ and define $m_t = m$.
If there is no such $u_i$, then continue to the next iteration.
\end{itemize}
It is useful to give an example of the step-by-step execution
of this algorithm, and 
Figures \ref{fig:alg-steps123b} and \ref{fig:alg-steps45b} (which
fill the next two pages) do this for a sample input.
In the notation of these figures, each language $L_i$ is a vertical column
with a cell for each string $u_j$, and an ``X'' in the cell indicates
that $u_j \in L_i$.
Each vertical column only goes up to the height $m_t$ in step $t$,
indicating that by the end of step $t$, the algorithm has only
considered finite prefixes of the form $L_i[m_t]$.

\agraf{h}{0.45}{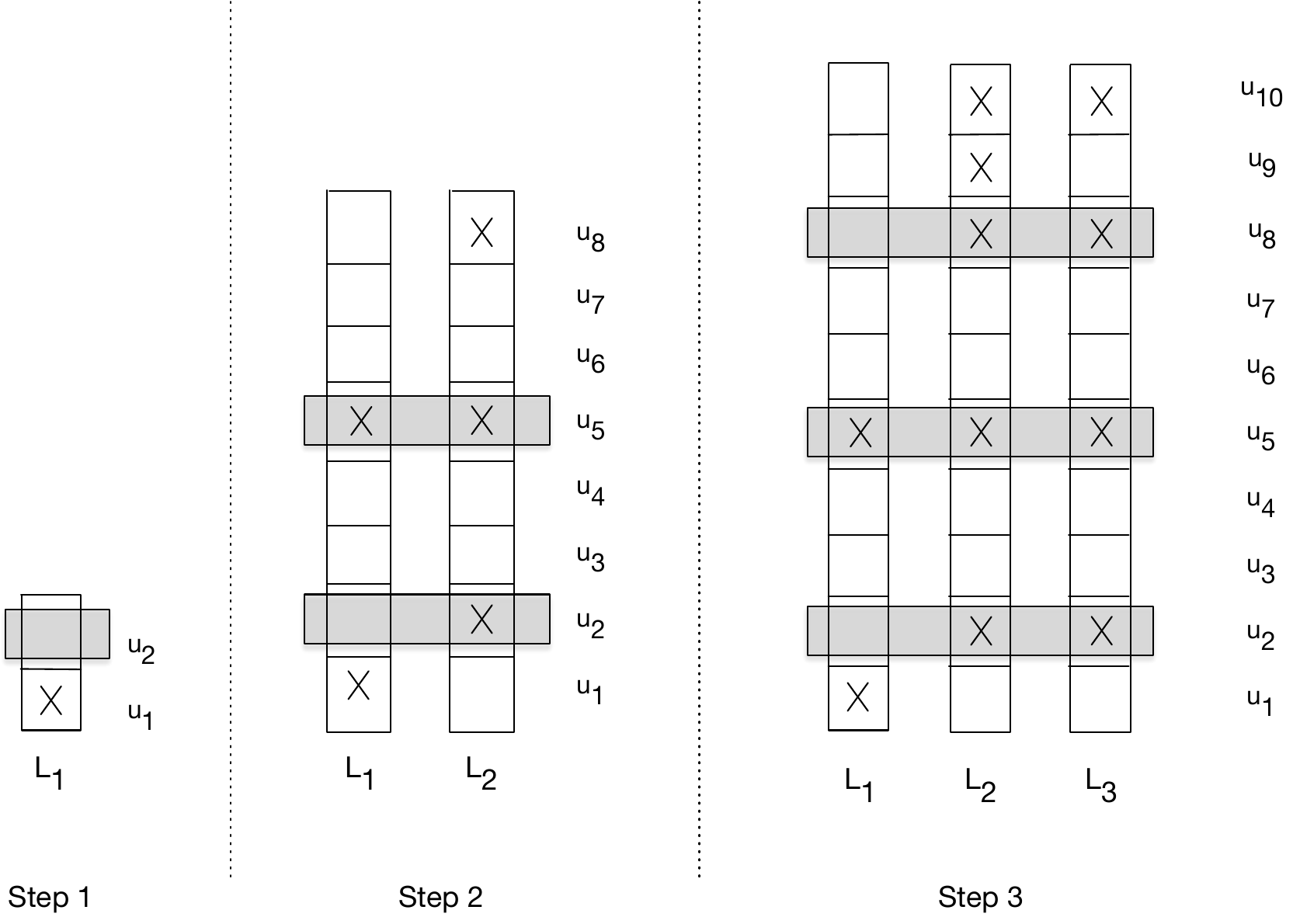}{fig:alg-steps123b}{
\small
This and the next figure show an example of the first five steps
of the algorithm from Section \ref{sec:gen-alg} on a sample input.
(It is useful to read the description of the algorithm before 
consulting this figure.)
The strings produced by the adversary in order over the first five
steps are $u_2, u_5, u_8, u_{10}, u_{12}$; 
in the notation of the figure, the steps of the algorithm are shown
separately, each language considered in a given step is shown as
a vertical column, and there is a row for each string considered at some
point in the step.  The string $u_j$ belongs to $L_i$ if and only if
there is an ``X'' in the column for $L_i$
and the row for $u_j$.  \\
\smallskip
~~ In step $t$, the rows corresponding to 
strings the adversary has already produced are shaded (so for example
in Step 2, the row for $u_8$ is not shaded because the adversary hasn't yet
produced $u_8$; it only does so in Step 3).  The algorithm considers
the languages in $\coll_t = \{L_1, L_2, \ldots, L_t\}$ 
in step $t$, and it only considers
a finite prefix of each language in step $t$, from an index
$\initmt$ at the start of the step to an index $m_t$ at the end. 
The ``heights'' of the colunms in each step go up to the final index $m_t$. \\
\smallskip
~~ Recall that the algorithm starts with $m = \initmt$, finds the
highest-indexed $(t,m)$-critical language $L_{n_t(m)}$ among $\coll_t$, and 
begins searching through strings of increasing index to find a new
string in $L_{n_t(m)}$.
During this search, as $m$ increases, the identity of the highest-indexed
$(t,m)$-critical language might change. \\
\smallskip
~~ In Step t = 1, there is no consistent language in $\coll_t$, so the 
algorithm can generate an arbitrary string.
Step t = 2 starts with $m = \initmt = 5$: 
$L_2$ is $(t,m)$-critical for all $m$, and so the algorithm
tests $u_m$ for membership in $L_2$ beginning at $m = 6$ until it finds
the first new string in $L_2$, which happens with $m = 7$;
$u_7$ is therefore generated.
Step t = 3 starts with $m = \initmt = 8$: 
$L_3$ is the highest-indexed $(t,8)$-critical language in $\coll_t$
(since $L_3[8] \subseteq L_2[8]$),
and so 
the algorithm begins searching for the next $u_m \in L_3$, as long as
$L_3$ remains $(t,m)$-critical; this happens at $m = 10$, and so 
$u_{10}$ is generated.}

\agraf{h}{0.58}{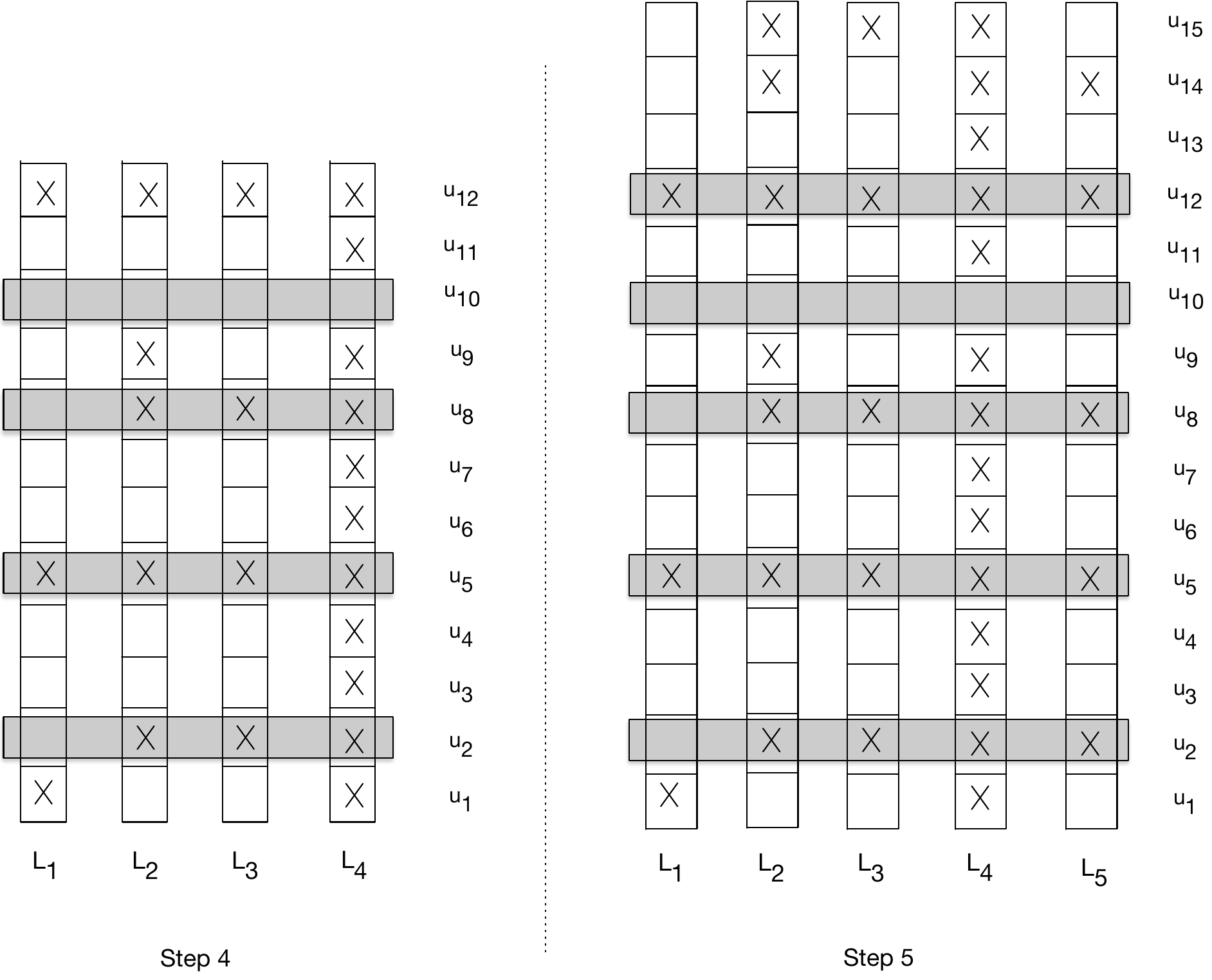}{fig:alg-steps45b}{
\small
This is a continuation of the execution of the algorithm on 
the sample input from Figure \ref{fig:alg-steps123b}.
Step t = 4 starts with $m = \initmt = 10$:, $L_4$ is consistent but not
$(t,10)$-critical (since $L_4[10] \not\subseteq L_3[10]$), 
and so $L_3$ remains the highest-indexed $(t,10)$-critical language.
As before,
the algorithm begins searching for the next $u_m \in L_3$, as long as
$L_3$ remains $(t,m)$-critical; this happens at $m = 12$, and so 
$u_{12}$ is generated. \\
\smallskip
~~ Step t = 5 starts with 
$m = \initmt = 12$: $L_5$ is the highest-indexed $(t,12)$-critical language
(since $L_5[12]$ is a subset of both $L_3[12]$ and $L_2[12]$), and so
the algorithm begins searching for the next $u_m \in L_5$, as long as
$L_5$ remains $(t,m)$-critical.
Once $m = 14$, however, the algorithm finds that $L_5$ is
not $(t,14)$-critical (since $L_5[14] \not\subseteq L_3[14]$),
and so $L_3$ is the highest-indexed $(t,14)$-critical language.
The algorithm switches to searching for the next $u_m \in L_3$, as long as
$L_3$ remains $(t,m)$-critical;
this happens when $m = 15$, and so $u_{15}$ is generated.
}

\xhdr{Analyzing the algorithm}
As written, it is not immediately clear that the algorithm's
iterations in step $t$ will come to an end with a string $u_m$,
rather than running forever.
We prove this now.

\begin{stmt}
In step $t$, the algorithm outputs a string after a finite sequence 
of operations.
\label{stmt:alg-terminates}
\end{stmt}

\proof{
We identify each iteration with the value of $m$ after 
the initial increment of the iteration;
so the iterations begin at $\initmt + 1$ and continue upward from there.
Suppose by way of contradiction that the algorithm performs an
infinite sequence of iterations.

Let us call an iteration $m$ {\em disruptive} if $n_t(m) \neq n_t(m-1)$.
Since $n_t(m-1)$ is the maximum index of a $(t,m-1)$-critical language,
and since our monotonicity property \rf{stmt:finite-critical-monotone}
implies that $L_{n_t(m)}$
is also $(t,m-1)$-critical, it follows
that $n_t(m) < n_t(m-1)$.
Since $n_t(m)$ starts at a value upper-bounded by $t$ and decreases
by at least one with every disruptive iteration, there can be at most
$t-1$ disruptive iterations.

The sequence of iterations must therefore 
contain a last disruptive iteration $m^*$.
For all iterations $m \geq m^*$, the language $L_{n_t(m)}$
does not change.
If there is an index $i \leq m^*$ for which 
\mbox{$u_i \in L_{n_t(m^*)} - \seen_t$}, 
then the algorithm terminates
in iteration $m^*$ with the first such $u_i$.
Otherwise, since the language $L_{n_t(m^*)}$
is infinite, we must eventually reach an iteration
$m > m^*$ for which $u_m \in L_{n_t(m)}$, and the algorithm
will stop and output $u_m$ at this point.
\endpf
}

\newpage 

Given that \rf{stmt:alg-terminates} establishes that 
the algorithm outputs a string in step $t$, it is useful to 
record an additional property of the algorithm that
follows directly from its construction.

\begin{stmt}
In step $t$, if at least one language in $\res{\coll}{t}$ is consistent with $\seen_t$,
then there is an $m_t$ and an $n_t$ such that 
the algorithm outputs a string from $L_{n_t(m_t)}[m_t]$, where $L_{n_t(m_t)}$ 
is the 
$(t,m_t)$-critical language with maximum index in $\res{\coll}{t}$.
\label{stmt:alg-max-critical}
\end{stmt}

We now prove an analogue of \rf{stmt:f-property}, from
which our main result \rf{stmt:main-infinite} directly follows.

\begin{stmt}
For any language $L_z \in \coll$ and any enumeration of $L_z$,
there is a $\fint$ such that for all $t \geq \fint$, the algorithm
generates a string in $L_z - \seen_t$.
\label{stmt:alg-property}
\end{stmt}

\proof{
In the given enumeration of $L_z$, 
\rf{stmt:finite-true-critical} tells us that 
there is a $\zt$ such that for all $t \geq \zt$ and all $m \geq 1$,
the language $L_z$ is $(t,m)$-critical.
Let $\fint = \max(z,\zt)$.
In every step $t \geq \fint$, 
by \rf{stmt:alg-max-critical} there is an $m_t$ such that
the algorithm generates a string from $L_{n_t(m_t)}[m_t]$, where
$L_{n_t(m_t)}$ is the $(t,m_t)$-critical language 
with maximum index in $\res{\coll}{t}$.
In each such step $t$, $L_z$ is a 
$(t,m_t)$-critical language in $\res{\coll}{t}$, and
so $n_t(m_t) \geq z$.
From \rf{stmt:finite-critical-nested}, it follows
that $L_{n_t(m_t)}[m_t] \subseteq L_z[m_t] \subseteq L_z$.
Since the algorithm's output comes from $L_{n_t(m_t)}[m_t] - \seen_t$,
it follows that it comes from $L_z - \seen_t$ as well.
\endpf
}

As in the discussion at the end of Section \ref{sec:gen-f},
it is straightforward to modify the algorithm so that it generates
strings without repetition.

\section{Generation for Finite Collections of Languages}
\label{sec:finite}

We now turn to our second main result, \rf{stmt:main-finite},
which derives a stronger conclusion for finite collections of languages.

The finite case illustrates the power of the
closure operator described in Section \ref{sec:obs}; 
this turns out to be sufficient to obtain the result.
To review the definition from Section \ref{sec:obs},
for a sequence of strings $\seen_t$ from a language in $\coll$,
we define the {\em closure} of $\seen_t$ in $\coll$, denoted
$\clos{\seen_t}$, to be the intersection of all languages
in $\coll$ that are consistent with $\seen_t$.
If there is a string in $\clos{\seen_t} - \seen_t$, then it is
always safe to generate such a string; by
definition, it must be an unseen string from the true language $L_z$.
The challenge in the previous sections was that there are simple
instances with infinite collections $\coll$ for which 
$\clos{\seen_t} = \seen_t$.
But for finite collections $\coll$, we will be able to make much more
progress using the closure operator.

\xhdr{Informal version of the argument}
We start by giving the basic idea behind the proof, and then 
the proof itself.
Let us write the finite collection of candidate languages as
$\coll = \{L_1, L_2, \ldots, L_n\}$, and suppose that
after the adversary has enumerated a set $\seen$ of strings, 
the languages consistent with $S$ are 
$L_{i_1}, L_{i_2}, \ldots, L_{i_k}$.
Note that the true language $\trueL$ must be one of these $k$ languages.
Now, the closure $\clos{S}$ is equal to the 
mutual intersection $L_{i_1} \cap L_{i_2} \cap \cdots \cap L_{i_k}$,
and there are two possibilities: either
$\clos{S}$ is infinite, or it is finite.
If it is infinite, then the algorithm can safely generate
all of the strings in $\clos{S} - S$, and thus achieve the goal
specified by \rf{stmt:main-finite}.
On the other hand, if $\clos{S}$ is finite, then it has size
equal to some natural number $m$;
in this case, after the adversary enumerates at most $m+1$ more distinct
strings, the algorithm will learn that at least one of 
$L_{i_1}, L_{i_2}, \ldots, L_{i_k}$ is no longer consistent.
We will then have a set of at most $k-1$ consistent languages, and we can
iterate this argument at most $k-2$ more times until 
(i) there is only a single consistent language, which must be $\trueL$, or
(ii) more generally, the set of all consistent languages has a 
mutual intersection that is infinite, in which case the algorithm can
safely generate from this infinite set.

This argument conveys the key point underlying the proof, that as the
adversary enumerates strings from $\trueL$, it cannot prevent itself
from reaching a point where the set of strings it has enumerated
has an infinite closure.
To turn this into an argument that produces a uniform bound on how
many strings are needed before the closure must become infinite,
we replace the iterative argument in the previous paragraph
with one that is shorter and more direct.
Specifically, consider all sub-collections of languages from $\coll$
(where we think of a sub-collection as any way of choosing some of
the languages from $\coll$ but not others).
Note that since $\coll$ is finite, there are only finitely
many possible sub-collections of languages from $\coll$.
For each sub-collection $\coll'$, the mutual intersection of the languages
in $\coll'$ is either finite or infinite.
Consider the sub-collections that have a finite mutual intersection,
and let $m^*$ be the maximum size of such a mutual intersection.
Now, suppose the adversary produces a set $\seen$ of $m^* + 1$
distinct strings from $\trueL$.
If we consider the sub-collection of all languages in $\coll$
that are consistent with $S$, its mutual intersection must contain $S$
and therefore it has cardinality greater than $m^*$.
By the definition of $m^*$, this means that its cardinality must be infinite.
So the closure $\clos{S}$ is infinite, and therefore the algorithm
can safely generate all the strings in $\clos{S} - S$.

\xhdr{The proof}
The argument above, in informal terms, 
is the complete proof of \rf{stmt:main-finite}.
We now formalize this argument in the remainder of the section.

\prevs{stmt:main-finite}{
We begin with some additional definitions.
For any subset $\indset$ of the indices $\{1, 2, \ldots, n\}$, let
$\allint(\indset)$ be the intersection of the languages whose indices are in $\indset$;
in other words,
$\displaystyle{\allint(\indset) = \bigcap_{i \in \indset} L_i.}$
For any sequence $\seen$ of strings from a language in $\coll$,
let $I(\seen)$ be the set of indices of the languages in $\coll$
that contain $\seen$;
that is,
$I(\seen) = \{i : \seen \subseteq L_i\}.$
We observe that the closure operator can be written in terms of this
notation, in that
$\clos{\seen} = \allint(I(\seen))$.

If $\allint = \allint(\{1, 2, \ldots, n\})$ is infinite, then 
the algorithm can generate arbitrary strings from $\allint$ as its output 
without seeing any sample of strings at all; since
$\allint \subseteq L_i$ for every language $L_i \in \coll$, 
in particular $\allint \subseteq L_z$ for the true language $L_z$, and
this satisfies the requirements of \rf{stmt:main-finite}.

For the rest of the proof, we therefore suppose 
$\allint(\{1, 2, \ldots, n\})$ is finite.
Let $\fincoll$ be the collection of all sets of indices 
$\indset \subseteq \{1, 2, \ldots, n\}$ 
with $\allint(\indset)$ finite.
Finally, let
$\displaystyle{m^* = \max_{\indset \in \fincoll} |D(A)|}$;
since $|\fincoll| \leq 2^n$, 
we observe that $m^*$ is the maximum of a finite set of positive integers,
and hence a positive integer.

We now define $t(\coll) = m^* + 1$ and claim that this choice
of $t(\coll)$ satisfies the required guarantee of \rf{stmt:main-finite}. 
Indeed, consider the true language $L_z \in \coll$ and any sequence 
$\seen_t$ of $t(\coll)$ distinct elements from $L_z$.
Recall that $I(\seen_t)$ denotes the 
set of indices of all languages in $\coll$ that contain $\seen_t$.
We have $\seen_t \subseteq \clos{\seen_t} = \allint(I(\seen_t))$.
If $\allint(I(\seen_t))$ were finite, then by the definition of $m^*$,
the cardinality of $\allint(I(\seen_t))$ would be at most $m^*$.
But this would contradict the fact that 
$\allint(I(\seen_t))$ contains $\seen_t$, which has cardinality $m^* + 1$.

Therefore $\clos{\seen_t} = \allint(I(\seen_t))$ is infinite,
and it is a subset of the true language $L_z$.
To conclude the proof, we therefore
need only show that there is an algorithm
that can enumerate all of $\clos{\seen_t} - \seen_t$ using only
membership queries.
To do this, the algorithm begins by querying whether each 
$w_i \in \seen_t$ belongs to each $L_j \in \coll$.
From this, it can determine the set $I(\seen_t)$ of indices
of languages that contain $\seen_t$.
Now, it enumerates every string $u_i \in U$ in ascending order,
skipping the elements of $\seen_t$.
For each such string $u_i$, it queries whether $u_i \in L_j$
for each $j \in I(\seen_t)$, and it outputs $u_i$ if it belongs
to each of these languages.
In this way, the algorithm enumerates the infinite set
$\clos{\seen_t} - \seen_t \subseteq L_z - \seen_t$ after
seeing a sample of $t(\coll)$ strings in $L_z$.
\endpf
}

\section{Extensions and Generalizations}
\label{sec:extend}

As discussed at the end of Section \ref{sec:model},
a natural direction for generalization is to consider
whether we can preserve the general structure of the model
while adding in a notion of {\em prompting} --- an idea
familiar from language generation systems in practice, where
the algorithm is provided with a prompt string and it must 
complete it to a valid output.

To explore how we might add
add prompts to our model,
let's first recall that 
there is a countable collection of
language $\coll = \{L_1, L_2, L_3, \ldots\}$, the 
adversary chooses a true language $\trueL$ from $\coll$,
and it begins enumerating the strings of $\trueL$ one by one, over
a sequence of steps $t = 1, 2, 3, \ldots$.
We continue to use $S_t$ to denote the set of strings 
enumerated by the adversary up through step $t$, and 
$z$ for an index such that $K = L_z$.

\xhdr{A model of prompting}
The new feature of the problem in our generalization is that
in every step $t$, the adversary provides the algorithm with two things:
a string $w_t$ from the true language $K$, and 
a string $p_t$ that serves as a {\em prompt}.  
(The adversary is allowed use the same prompt in multiple steps.)
The algorithm in step $t$ must then produce a string $c_t$ with
the goal that the concatenation of $p_t$ and $c_t$ is a string belonging 
to $\trueL - S_t$, where
$S_t = \{w_1, w_2, \ldots, w_t\}$ --- 
that is, it should be an unseen string from $\trueL$.
In what follows, we will use $p_t \cat c_t$ to denote the
concatenation of $p_t$ and $c_t$.

We observe that it leads to an equivalent problem whether we 
ask the algorithm to output $c_t$ so that $p_t \cdot c_t \in \trueL - S_t$,
or whether we ask the algorithm to output the full contatenated string
$a_t = p_t \cat c_t$.
In this latter formulation, we can phrase the algorithm's task as
follows: 
given a prompt $p_t$, output a string $a_t$ with the properties that
$p_t$ is a prefix of $a_t$, and $a_t \in \trueL - S_t$.
Because it makes the exposition slightly simpler, 
we will take this latter formulation as our default version of the
problem --- that the algorithm is supposed to output the full
string $a_t$, with $p_t$ as a prefix of $a_t$ ---
but we will refer to both versions in our discussion.

To establish a positive result for prompted generation in the limit,
we need
to impose some type of restriction on the prompts the adversary
can provide.
For example, if the adversary were simply to provide an arbitrary
string $p_t$ and ask the algorithm if there exists a 
string $c_t$ and a language $L_i \in \coll$ for which 
$p_t \cdot c_t \in L_i$, this is not a problem that could 
be solved by an algorithm that must terminate with a yes/no answer and 
whose only access to $\coll$ comes in the form of
membership queries of the form ``Is $w \in L_i$?''
So as a minimal restriction on the adversary, we can at least
require that its prompt $p_t$ at step $t$ must have the
property that there exists a string $c_t$ for which
$p_t \cat c_t \in \trueL - S_t$.
This weak assumption raises interesting open questions that we
will consider later in this section.
But first, we will establish a positive result with a stronger
restriction on the adversary, as follows.
We say that a prompt $p$ is {\em robust} if for all 
languages $L_i \in \coll$, there exist arbitrarily long strings $c$
for which $p \cat c \in L_i$.
We will start by considering adversaries that only provide robust prompts.

We say that the algorithm achieves {\em prompted generation from 
$\trueL$ in the limit} if there is some $t^*$ such that 
for all steps $t \geq t^*$, the algorithm's output $a_t$
has the property that $p_t$ is a prefix of $a_t$ and $a_t \in \trueL - S_t$.
We now prove the following.

\begin{stmt}
There is an algorithm with the property that for any countable
collection of languages $\coll = \{L_1, L_2, L_3, \ldots\}$,
and any enumeration of one of these languages $\trueL$ accompanied
by a sequence of robust prompts,
the algorithm achieves prompted generation from $\trueL$ in the limit.
\label{stmt:main-prompted}
\end{stmt}

We make two initial observations about this result.
First, \rf{stmt:main-prompted} is a strict
generalization of our first main result
\rf{stmt:main-infinite}, since if the adversary always provides
the empty string as its prompt $p_t$, then the problem of finding
continuations $c_t$ for which $p_t \cat c_t \in \trueL - S_t$ is
simply the problem of finding strings $c_t$ in $\trueL - S_t$,
as in the original definition of generation in the limit.
Moreover, the empty string is a robust prompt, since
each of the languages $L_i \in \coll$ is infinite, 
and so there are arbitrarily
long continuation strings that belong to $L_i$ when
concatenated to the empty string.

Second, we observe that there is no requirement that the algorithm
has ever seen a string beginning with the prefix $p_t$ among
the adversary's examples $S_t \subseteq \trueL$ before
the first step $t$ in which the adversary provides $p_t$.
An important point to notice about 
\rf{stmt:main-prompted}
is that the algorithm can achieve prompted
generation in the limit despite this challenge.

\subsection{A First Result for Prompted Generation}
\label{sec:robust-prompt}

We now describe how to prove our result \rf{stmt:main-prompted}.
The proof is a direct adaptation of the proof of 
\rf{stmt:main-infinite} from Section \ref{sec:gen-alg};
as we will see, the structure of critical languages built up there
is sufficiently strong that not much more is needed to handle
the prompted version of the problem with robust prompts.

As in Section \ref{sec:gen-alg}, we will work with a specific
enumeration $u_1, u_2, u_3, \ldots$ of all strings in $U$,
and work with finite subsets of the languages $L_i$,
defined via the notation $L_i[m] = L_i \cap \{u_1, u_2, \ldots, u_m\}$.
The algorithm for prompted generation will closely
follow the algorithm from Section \ref{sec:gen-alg},
in that in every step $t$, it will increment a counter $m$
and maintain knowledge of the maximum index $n_t(m)$
of a $(t,m)$-critical language
from $\coll_t$.
Maintaining knowledge of $n_t(m)$ does not require knowledge
of the prompts, and so this part of the algorithm is the same as before.
What changes is the stopping condition for the algorithm in step $t$:
rather than continue increasing $m$ until any valid output is found ---
that is, until $u_m \in L_{n_t(m)}$ --- the
algorithm must increase $m$ potentially even further, until it finds
a string $u_m$ for which $p_t$ is a prefix of $u_m$, and
$u_m \in L_{n_t(m)} - S_t$.
However, since $p_t$ is a robust prompt, 
the algorithm is guaranteed to eventually find such a string,
and so we can be sure that its iterations in step $t$ will terminate.
If we let $m_t$ be the value of $m$ at the end of step $t$, then 
once $t$ is large enough, we know that 
$L_{n_t(m_t)}[m_t] \subseteq L_z[m_t]$, 
where $L_z = \trueL$ is the true language, and
so the string $u_m$ that it outputs has $p_t$ as a prefix and belongs to 
$\trueL - S_t$ as required.

\xhdr{Detailed analysis}
The discussion above probvides the entire set of modifications 
to the algorithm; 
for completeness we now describe these in more detail, together
with a proof of correctness.

First, the facts
\rf{stmt:finite-true-critical} through 
\rf{stmt:finite-critical-monotone} still hold in the prompted
case, since they are structural properties of the language
that are not affected by the adversary's use of prompts.
The algorithm for generating an output string uses an iteration
in step $t$ for which parts (i), (ii), and (iii) of each iteration
are the same as in Section \ref{sec:gen-alg}.
Step (iv) of each iteration is replaced by
\begin{itemize}
\item[(iv$'$)] If there is any string $u_i$ for $i \leq m$ such that
$u_i$ has $p_t$ as a prefix and
$u_i \in L_{n_t(m)} - S_t$, then 
choose the minimum $i$ with this property;
output the string $u_i$ and define $m_t = m$.
If there is no such $u_i$, then continue to the next iteration.
\end{itemize}

Now, the proof of termination works as before, by establishing
that there are only finitely many {\em disruptive iterations} in which
the identity of $n_t(m)$ changes;
this part does not depend on the structure of prompts but
only on the definition of a $(t,m)$-critical language, and so it uses 
\rf{stmt:finite-true-critical} through
\rf{stmt:finite-critical-monotone} exactly as before.
After the last disruptive iteration, either 
there is a string $u_i \in L_{n_t(m)} - S_t$ with $i \leq m$ 
for which $p_t$ is a prefix, or else the algorithm will eventually reach one,
since the prompt $p_t$ is robust.
It declares this $u_m$ to be its output string.
We therefore have

\begin{stmt}
In step $t$, if at least one language in $\res{\coll}{t}$ is consistent with $\seen_t$,
then there is an $m_t$ and an $n_t$ such that 
the algorithm terminates with 
a string $a_t$ for which $p_t$ is a prefix of $a_t$ and 
$a_t \in L_{n_t(m_t)}[m_t] - S_t$, where $L_{n_t(m_t)}$ is the 
$(t,m_t)$-critical language with maximum index in $\res{\coll}{t}$.
\label{stmt:prompted-max-critical}
\end{stmt}

Finally, we establish the basic correctness property of the algorithm,
from which \rf{stmt:main-prompted} follows directly.

\begin{stmt}
For any language $L_z \in \coll$ and any enumeration of $L_z$
with robust prompts $p_1, p_2, p_3, \ldots$, 
there is a $\fint$ such that for all $t \geq \fint$, the algorithm
generates a string $a_t$ for which $p_t$ is a prefix of $a_t$
and $a_t \in L_z - \seen_t$.
\label{stmt:prompted-property}
\end{stmt}

\proof{
In the given enumeration of $L_z$, 
\rf{stmt:finite-true-critical} tells us that 
there is a $\zt$ such that for all $t \geq \zt$ and all $m \geq 1$,
the language $L_z$ is $(t,m)$-critical.
Let $\fint = \max(z,\zt)$.
In every step $t \geq \fint$, 
by \rf{stmt:prompted-max-critical} there is an $m_t$ such that
the algorithm generates 
a string $a_t$ such that
$p_t$ is a prefix of $a_t$, and 
$a_t \in L_{n_t(m_t)}[m_t] - S_t$, where $L_{n_t(m_t)}$ is the 
$(t,m_t)$-critical language with maximum index in $\res{\coll}{t}$.
In each such step $t$, $L_z$ is a 
$(t,m_t)$-critical language in $\res{\coll}{t}$, and
so $n_t(m_t) \geq z$.
From \rf{stmt:finite-critical-nested}, it follows
that $L_{n_t(m_t)}[m_t] \subseteq L_z[m_t] \subseteq L_z$.
Since $a_t \in L_{n_t(m_t)}[m_t] - \seen_t$,
it follows that $a_t \in L_z - \seen_t$ as well.
\endpf
}

\subsection{Prompted Generation with a More Powerful Adversary}

Having proved \rf{stmt:main-prompted},
let us go back to the question of what we must require about the
adversary's prompts in order to establish a positive result for
prompted generation.
Arguably the weakest requirement we might picture placing on the 
adversary --- and therefore, the most power we could give the adversary ---
would be to require only
that each of its prompts has at least one valid continuation
at the time that it poses the prompt:
that is, its prompt $p_t$ in step $t$ must have the property that
there is at least one string $c_t$ for which $p_t \cat c_t \in \trueL - S_t$.
Let us call such a prompt {\em non-trivial}.
While requiring non-trivial prompts is a 
weaker restriction than requiring robust prompts as 
in Section \ref{sec:robust-prompt}, 
we reiterate the point that even the assumption of robust prompts
is sufficient to provide a strict generalization of our first
main result \rf{stmt:main-infinite}, given that the empty string,
when used as a prompt, is robust.

We now establish some results with this weaker restriction on the
adversary, though we 
leave a fully general characterization of the power of this
restriction as an open question.
In particular, going back to the style of results from
Section \ref{sec:gen-f}, where we allow algorithms with additional
computational power, we will show that there is a positive
result for non-trivial prompts via an algorithm that can ask
not only membership queries of the form 
``Is $w \in L_i$?'' but is also 
augmented with one extra form of computational power: 
if $L_i \in \coll$ and $R$ is a regular language (i.e. one
that is accepted by a finite automaton), then 
the algorithm can correctly answer the 
query ``Is $L_i \subseteq R$?''.\footnote{Equivalently, 
we could say that the algorithm's added power is to answer 
questions of the form ``Is $L_i \cap R'$ empty?'' for a
regular language $R'$, since $R$ is a regular language if and only
if its complement $U - R$ is, and $L_i$ is a subset of $R$
if and only if $L_i$ is disjoint from $U - R$.}
The fact that such an algorithm can perform prompted generation
in the limit for any sequence of non-trivial prompts has
an interesting implication: we can establish this same result
without any augmentation of the algorithm by additional power --- that is, using
an algorithm that can only perform standard computational steps ---
in the case where $\coll$ is the set of all context-free languages.
This is simply because the problem ``Is $L \subseteq R$?'' is
decidable when $L$ is context-free and $R$ is regular, and
so we don't need to augment the algorithm with any additional
power to answer these queries when $\coll$ consists of
the set of context-free languages.\footnote{To see why this is
decidable, we begin with the point in the previous footnote,
that for the algorithm to answer ``Is $L \subseteq R$?'' it can
equivalently answer ``Is $L \cap (U - R)$ empty?''
The intersection of a context-free language with a regular language
is context-free, and the question of whether a context-free language is
empty is decidable.}

\xhdr{Prompted generation using regular subset queries}
To make the discussion above concrete, we write the added
computational power of the algorithm as the following assumption.

\begin{stmt}
Assumption: The algorithm is able to answer subset queries of the form
``Is $L_i \subseteq R$?'' where $L_i \in \coll$ and $R$ is 
a regular language.
\label{stmt:subset-assumption}
\end{stmt}

We will call the type of query specified in 
\rf{stmt:subset-assumption} a {\em regular subset query},
and again, we note that for some important families of languages ---
for example the context-free languages ---
an algorithm does not need to be
augmented with anything additional in order to decide
regular subset queries.

We will prove the following.

\begin{stmt}
There is an algorithm, augmented with the power to perform
regular subset queries of the form in Assumption \rf{stmt:subset-assumption},
with the property that for any countable
collection of languages $\coll = \{L_1, L_2, L_3, \ldots\}$,
and any enumeration of one of these languages $\trueL$ accompanied
by a sequence of non-trivial prompts,
the algorithm achieves prompted generation from $\trueL$ in the limit.
\label{stmt:non-trivial-prompted}
\end{stmt}

As in Subsection \ref{sec:robust-prompt}, we prove this 
by adapting the algorithm from Section \ref{sec:gen-alg}.
However, this time the modifications need to be a bit more extensive.
As a first new definition, 
we call a language $L_i \in \coll$ 
{\em $t$-valid} if there exists a string $c$ such that
$p_t \cat c \in L_i - S_t$.
Observe that $\trueL = L_z$ is $t$-valid for all $t$,
by our requirement that the adversary has to provide non-trivial prompts.

\xhdr{An algorithm for generation with non-trivial prompts}
We now give the algorithm that proves \rf{stmt:non-trivial-prompted}.
As in Section \ref{sec:gen-alg}, in step $t$
we maintain a prefix size $m_t$ and work with
prefixes $L_i[m_t]$ for $i \leq t$.
As before, we write the sequence of examples provided so far 
by the adversary as $S_t = \{w_1, w_2, \ldots, w_t\}$.

We let $n_t(m)$ be the maximum index of a $(t,m)$-critical language
in $\coll_t$, if such an index exists.
We let $r_t(m)$ be the maximum index of a language in $\coll_t$
that is both $(t,m)$-critical and $t$-valid, if such an index exists.
We continue to use \rf{stmt:finite-true-critical}, which
establishes that there is some time step $\zt$ such that for all
$t \geq \zt$, and all $m \geq 1$, the language 
$L_z$ is $(t,m)$-critical.
If we set $t^* = \max(z,\zt)$, then for all $t \geq t^*$,
the quantities $n_t(m)$ and $r_t(m)$ are both well-defined for all $m$, 
since when $t \geq t^*$,
there is at least one language in 
$\res{\coll}{t}$ (namely $L_z$) that is both 
$(t,m)$-critical for all $m$ and also $t$-valid.

In step $t$, the algorithm is looking for a string $u \in U - S_t$
to output that has $p_t$ as a prefix;
we will call this an {\em acceptable} output.
As a first phase of step $t$, the algorithm determines which
languages in $L_i$ are $t$-valid.
To do this, it defines $R_t$ to be the language consisting of all
strings that have $p_t$ as a prefix, and it then
checks whether $L_i \cap (R_t - S_t)$ is empty.
Since $R_t - S_t$ consists of all acceptable outputs,
this check determines whether or not $L_i$ is $t$-valid.
Moreover, since $R_t - S_t$ is a regular language,
this check can be done using the algorithm's augmented
power to determine whether $L_i$ is a subset of a given regular language,
or equivalently whether $L_i$ has an empty intersection with a given
regular language.

Operating by analogy with 
Section \ref{sec:gen-alg}, the algorithm then defines a counter $m$
that it initalizes to $\max(m_{t-1},w_t)$, where $w_t$ is the
most recent string in $S_t$.
If $\coll_t$ fails to contain a language
that is both $(t,m)$-critical and $t$-valid, then
the algorithm can output an arbitrary string in step $t$.
(This will not pose a problem, since as we have observed above
\rf{stmt:finite-true-critical} establishes that $\coll_t$ will always
contain such languages once $t$ is large enough.)
Henceforth, let us assume we are at a step $t$ for which 
$\coll_t$ contains at least one language 
that is both $(t,m)$-critical and $t$-valid, which means that
$n_t(m)$ and $r_t(m)$ are well-defined. 
In this case, the 
algorithm performs a sequence of iterations to find a string to output,
where each iteration consists of the following steps.
\begin{itemize}
\item[(i)] Increment $m$ by 1.
\item[(ii)] Perform membership queries to determine $L_i[m]$ for each 
$L_i \in \res{\coll}{t}$.
Note that since $m > \max_{t' < t} w_{t'}$, 
the determination of which languages in $\res{\coll}{t}$ are consistent
with $\seen_t$ does not change 
(relative to the initial iteration) when we do this.
\item[(iii)] Determine which languages are $(t,m)$-critical, and from 
this determine $n_t(m)$ and $r_t(m)$.
Note that this only requires consulting the results of
membership queries already performed, together with earlier determination
of which languages $L_i \in \coll_t$ are $t$-valid.
\item[(iv)] If there is any string $u_i$ for $i \leq m$ such that
$u_i$ has $p_t$ as a prefix and $u_i \in L_{r_t(m)} - S_t$, then 
choose the minimum $i$ with this property;
output the string $u_i$ and define $m_t = m$.
If there is no such $u_i$, then continue to the next iteration.
\end{itemize}

The proof that the iterations in step $t$ terminate with an output string
again follows the general structure of the proof of 
\rf{stmt:alg-terminates}.
In particular, we say that an iteration $m$ is {\em disruptive}
if $r_t(m) \neq r_t(m-1)$.
Since \rf{stmt:finite-critical-monotone} implies that
the indices of $(t,m)$-critical languages form a subset of
the indices of $(t,m-1)$-critical languages, and since 
the identities of the $t$-valid languages do not depend on the value of $m$,
we have $r_t(m) < r_t(m-1)$ after a disruptive iteration, and
therefore there can be at most $t-1$ disruptive iterations.
Now, consider the final disruptive iteration $m$.
Step (iv) of this iteration $m$ checks whether any
string among $u_1, \ldots, u_m$ has $p_t$ as a prefix and
belongs to $L_{r_t(m)} - S_t$, and if so
the iterations terminate with the earliest among these strings as output.
If none of $u_1, \ldots, u_m$ has this property, then 
the fact that $L_{r_t(m)}$ is $t$-valid implies that there is some
$u_{m'}$ with $m' > m$ that will have this property.
Therefore, by the time iteration $m'$ is reached, the algorithm will
terminate with an acceptable output.

The following claim now establishes 
\rf{stmt:non-trivial-prompted}.

\begin{stmt}
For any language $L_z \in \coll$ and any enumeration of $L_z$
with non-trivial prompts $p_1, p_2, p_3, \ldots$, 
there is a $\fint$ such that for all $t \geq \fint$, the algorithm
generates a string $a_t$ for which $p_t$ is a prefix of $a_t$
and $a_t \in L_z - \seen_t$.
\label{stmt:non-trivial-prompted-property}
\end{stmt}

\proof{
In the given enumeration of $L_z$, 
\rf{stmt:finite-true-critical} tells us that 
there is a $\zt$ such that for all $t \geq \zt$ and all $m \geq 1$,
the language $L_z$ is $(t,m)$-critical.
Let $\fint = \max(z,\zt)$.
As argued above, $r_t(m)$ is well-defined 
in every step $t \geq \fint$, 
and the algorithm generates a string $a_t$ such that
$p_t$ is a prefix of $a_t$ and
$a_t \in L_{r_t(m_t)}[m_t] - S_t$.
In each such step $t$, $L_z$ is a 
language in $\res{\coll}{t}$ that is both $(t,m_t)$-critical and $t$-valid,
and so $r_t(m_t) \geq z$.
From \rf{stmt:finite-critical-nested}, it follows
that $L_{r_t(m_t)}[m_t] \subseteq L_z[m_t] \subseteq L_z$.
Since $a_t \in L_{r_t(m_t)}[m_t] - \seen_t$,
it follows that $a_t \in L_z - \seen_t$ as well.
\endpf
}

\subsection{Prompted Generation for a Finite Collection of Languages}

When the collection $\coll$ of languages is finite, we can ask a final
question, which is whether an analogue of our result
\rf{stmt:main-finite} might hold for prompted generation:
could there be an algorithm, together with
some fixed bound $t(\coll)$, such that after
seeing $t(\coll)$ strings from the true language $\trueL$,
the algorithm is able to achieve prompted generation?

There is a short argument that such a result cannot hold.
To see why, suppose that $\coll$ is the collection of two languages
$L_1$ and $L_2$, each over the two-letter alphabet $\{a, b\}$,
where $L_1$ consists of all strings that begin with $a$ 
and all odd-length strings that begin with $b$;
and $L_2$ consists of all strings that begin with $a$ 
and all even-length strings that begin with $b$.
Suppose there were a bound $t(\coll)$ and an algorithm that guaranteed
correct prompted generation after seeing $t(\coll)$ distinct samples
from the true language $\trueL$.
Then an adversary could present $t(\coll)$ distinct strings 
all beginning with $a$, and then provide the single-letter string $b$
as a prompt:
if the algorithm outputs an even-length string, then the adversary
could declare this to be incorrect because the true language is $L_1$,
and conversely if the algorithm outputs an odd-length string.

This does not prevent the algorithm from achieving promoted
generation in the limit, because the adversary must eventually
output a string beginning with $b$, after which 
the algorithm knows whether to respond to prompts that begin with $b$
using even-length or odd-length strings.
But this guarantee cannot be achieved after any fixed number
of samples $t(\coll)$ that must be specified in advance.

\section{Concluding Remarks}
\label{sec:final}

Our results suggest that 
generating from a language based on observed samples is
a fundamentally different, more tractable problem than identifying
the language from observed samples.
It is so tractable, in fact, that it can be accomplished
provided only that we are told the samples come from a language
in a known countable collection of candidate languages.

It is therefore interesting to ask what stylized conclusions 
we might draw from these general results
about generation as a task, and its relation to other learning processes.
In the case of finite collections of languages, the basic idea
underlying the proof is that a large ``core'' to the language
(the closure of the sample, in our terminology) 
emerges at a known time after a finite set of observations, and it is then
enough to generate from this core even though there might always
remain peripheral parts of the language --- disjoint from this
core --- that we can never be sure about.
In the case of infinite collections of languages, the task is
more complex, because there is never a known time at which
a core to the language emerges.
Instead, the algorithm may need to continually shrink
the set it is using for generation;
through this infinite process
of shrinkage, the algorithm can be sure that beyond a certain point,
it is always generating from the true language $\trueL$, even if it can
not be sure when it has reached this point or what the true language is.

In this way, as noted earlier in the paper, the solutions we develop highlight 
some interesting tensions between the problem of producing {\em valid}
strings that belong to the target language, and the problem of maintaining
{\em breadth} by not restricting to only a limited subset of the target language.
Our approaches achieve validity through a strategy that implicitly
gives up on breadth as a goal, and it is interesting to ask whether
this is essentially necessary for any method that achieves
language generation in the limit.

This tension, as it arises in our solution, also creates an interesting
echo of the human process by which people acquire the vernacular
within a new community \cite{danescu2013no}:
as with our solution in this abstract model, people encountering
the dialect in a new community may 
similarly pass through a sequence of
conceptually distinct phases:
an initial phase in which they are generating too 
adventurously and producing invalid utterances;
then a phase where the utterances are approximately 
aligned with the scope of the language; 
and finally a phase in which 
the range of utterances they are willing to generate
shrinks further over their lifetime, as they become increasingly
conservative in what they consider valid.
Again, 
it is interesting to consider whether this type of structure is
inherent in any solution to the task of generation in the limit.

\xhdr{Acknowledgements}
We thank Bobby Kleinberg, Lillian Lee, Marios Papachristou,
and Kenny Peng for helpful discussions on these questions
and on early drafts of this paper.
The work has been 
supported in part by 
a Vannevar Bush Faculty Fellowship,
a Simons Collaboration grant,
a grant from the MacArthur Foundation,
and the Center for Applied AI at the University of Chicago
Booth School of Business.

\end{document}